\documentclass[twocolumn,showpacs,preprintnumbers,amsmath,amssymb,prl]{revtex4}

\usepackage{graphicx}% Include figure files
\usepackage{bm}% bold math

\begin{document}
\def\av#1{\langle#1\rangle}

\preprint{}

\title{Stability  and topology of scale-free networks under  attack and defense strategies}
\author{Lazaros K. Gallos$^1$} \author{Reuven Cohen$^2$}
\author{Panos Argyrakis$^1$} \author{Armin Bunde$^3$}
\author{Shlomo Havlin$^2$}
\affiliation{$^1$Department of Physics, University of Thessaloniki, 54124 Thessaloniki, Greece}
\affiliation{$^2$Minerva Center and Department of Physics, Bar-Ilan University, 52900 Ramat-Gan, Israel}
\affiliation{$^3$Institut f\"{u}r Theoretische Physik III, Justus-Liebig-Universit\"{a}t Giessen, Heinrich-Buff-Ring 16, 35392 Giessen, Germany}

\date{\today}% It is always \today, today,

\begin{abstract}
We study tolerance and topology of random scale-free networks
under  attack and defense strategies that depend on 
the degree $k$ of the nodes. This situation occurs, for example,
when the robustness of a node depends on its degree or in an intentional
 attack with insufficient knowledge on the network.
We determine, for all strategies, the critical fraction $p_c$ of nodes that must be removed  for
disintegrating the network. We find that for an intentional attack,  little knowledge  of the well-connected sites
is sufficient to strongly reduce $p_c$. At criticality, the topology
of the network depends on the  removal strategy, implying that different strategies  may lead to different kinds of percolation transitions.
\end{abstract}


\maketitle

The observation that many real networks, such as the Internet, the WWW,
social and biological networks etc, obey a power law distribution in
their nodes connectivity has inspired a  new area of research
\cite{AB02, DM02, PastorXX, ne, BA99, AB99, KRL00, Lil01}. Such a network is
constructed by nodes connected with links, where the probability $P(k)$ that a node
has $k$ links is
\begin{equation}
P(k) \sim k^{-\gamma} \, ,
\end{equation}
where $\gamma$ is usually found to be between $2<\gamma <3$.

The scale-free character of these networks, represented by having  no characteristic number of nodes per link, gives rise to many 
different and usually unexpected results in many properties, as compared
to lattice models or even to small-world networks \cite{AB02}.
One important feature studied is the robustness of such a network under
 a random node removal \cite{AJB00}. In general, the integrity of a network is destroyed after a critical percentage $p_c$ of the
 system nodes has been removed. For scale-free networks it has been shown that $p_c=1$, i.e., in order to destroy the network
 practically all nodes have to be removed \cite{Cohen00}.

In this Letter, we consider scale-free networks where the robustness of a node depends on 
its connectivity. This means that the probability of damaging a node either by some kind of failure
or by an external attack depends on the degree $k$ of the node. Examples are computer networks and social networks.
In a computer network such as the Internet, usually the hubs that serve many computers are built in a more robust way, so that
their probability of failure is smaller than for the others. In a social network, members 
of a group that have more links to others have a lower probability of leaving the group. It is, however, 
also possible that nodes with more links are less robust. For example, traffic on a network induces 
high loads on highly connected nodes \cite{weighted}, which in turn makes them more vulnerable to failures.
In some cases breakdowns are due to cascades of failures caused by the dynamics
of damage spreading~\cite{motter}. In computer networks many breakdowns are due
to congestion building (see~\cite{congestion}).

The above examples represent {\it internal} properties of a network, where the vulnerability of each
node depends on its degree.  In addition, the probability of removing a node can depend on its degree also
due to {\it external} attack strategy. For example, the most efficient attack is an intentional attack where the
highest degree nodes are being removed with probability one. In this case, only a small fraction $p$ of 
removed nodes is sufficient to destroy the network. This strategy, however, requires full knowledge of the 
network topology in order to identify the highest connected nodes. In many realistic cases, this information
is not available, and only partial knowledge exists. Accordingly, in an intentional attack the high-degree nodes
can be removed only with a certain probability that will depend on $k$.
 In some networks, such as Terror or Mafia networks those that are higher in the hierarchy have more
links and are less known, and therefore the probability to remove them is smaller than those with less links.
In contrast, in normal social networks the situation is opposite: The better linked members are more visible
and therefore have a higher probability to be attacked.
 Finally, our study also applies to
immunization strategies, where the high degree nodes are not always known in 
advance,
see also Dezs\"o and
Barab\'asi~\cite{DB02}.

In all these examples, a value $W(k_i)$ is assigned to each node,
which represents the probability that a node $i$ with $k_i$ links becomes inactive either by failure or under some attack. Specifically,
we focus on the family of functions,
\begin{equation}
\label{Wki}
W(k_{i}) = \frac{k_{i}^\alpha}{ \sum_{i=1}^{N} k_i^\alpha} \, ,\, 
-\infty<\alpha<\infty, 
\end{equation}
The parameter $\alpha$ can be the sum of two parameters, $\alpha=\alpha' + \alpha''$,
which incorporates both intrinsic network vulnerability ($\alpha'$) and
external knowledge of the system ($\alpha''$). It is possible that although some highly connected
nodes are intrinsically vulnerable ($\alpha'>0$), an internal defense strategy with $\alpha''<0$ will give the net result $\alpha<0$ \cite{footnote}. In this case,
   nodes with lower $k$  are more
vulnerable, while for $\alpha>0$, nodes with larger $k$ are more vulnerable.
 The cases $\alpha=0$ and $\alpha \to \infty$ 
represent the known random
removal \cite{AJB00,Cohen00} and the targeted intentional attack 
\cite{AJB00,Cohen00}, respectively.

The choice of $\alpha$ can be related to the probability $w$ that in each attack, one of the $n$  highest connected nodes in a network of $N$ nodes becomes inactive by failure
or attack.
By definition
\begin{equation}
\label{wdef}
w = \frac{\int_{k_n}^{k_{\rm max}} P(k) W(k) dk}{\int_{m}^{k_{\rm max}} P(k) W(k) dk} \,,
\end{equation}
where $m$ denotes the minimum number and $k_{\rm max}\sim N^{1/(\gamma-1)}$ the maximum
number of links a node can have. $k_n$ is the
minimum number of links of a node that belongs to  the $n$ highest connected nodes.
It is easy to verify that for $N \gg n \gg 1$, $w$ is related to $\alpha$ and $\gamma$ by
\begin{equation}
w = \frac{1-n^{(\gamma-1-\alpha)/(\gamma-1)} m^{\alpha+1-\gamma}}
{1-N^{(\gamma-1-\alpha)/(\gamma-1)} m^{\alpha+1-\gamma}}
 \,
\end{equation}
Depending on the value
of $\alpha$, for fixed value of $\gamma$, we can distinguish 3 different regimes.

\noindent
(i)  $\alpha > \gamma-1$: Here, $w=1$, and  an ``attacker'' is capable
of destroying all the highest nodes in the network. Even though $\alpha$ is finite here, this case is fully equivalent to the targeted
intentional attack $\alpha\to\infty$.

\noindent
(ii)  $\alpha= \gamma-1$: For this particular value of $\alpha$, $w$ depends logarithmically on $n$ and $N$,  $w=\ln n/\ln N$.

\noindent
(iii)  $\alpha< \gamma-1$: Here, $w$ decreases with $n/N$ by a power law,
$w=(n/N)^{1-\alpha/(\gamma-1)}$.
A special case is $\alpha=0$, where effectively nodes are picked randomly and
thus $w=n/N$. In this case, it is difficult to destroy the network, and the percolation threshold $p_c$ can even be $p_c=1$ for well-connected networks ($\gamma<3$)
\cite{AJB00,Cohen00}.
 For $0<\alpha\le \gamma-1$, the effective removal is better than random;
 $w$ still approaches zero for $N$ approaching infinity, but less fast than in the random case. As is shown later,
this is enough for the percolation threshold to be significantly smaller than 1.
Finally, for negative values of $\alpha$,
 the fraction $w$ of the highest 
connected nodes that will be damaged decreases much faster than in the random case. This feature will have no effect on the percolation threshold for $2<\gamma\le 3$,
where $p_c=1$ already in the random case. But for $\gamma>3$, $p_c$ increases with
increasing negative values of $\alpha$, as we show below.

From the above discussion it is clear that the regime $\alpha > \gamma-1$ is in the same universality class as the intentional
attack, and $\alpha=0$ is identical to the random attack. It is not clear if there are more
universality classes and if not, what is the border line between them.  To study this question, we have considered both numerical simulations and analytical
considerations. First we determine the percolation thresholds as a function of the network parameters $\gamma$, $N$ and $m$,  and  on the `attack' parameter
$\alpha$. Then we study the universality classes
by analyzing the topology of the network
at these thresholds.

\begin{figure}
\includegraphics{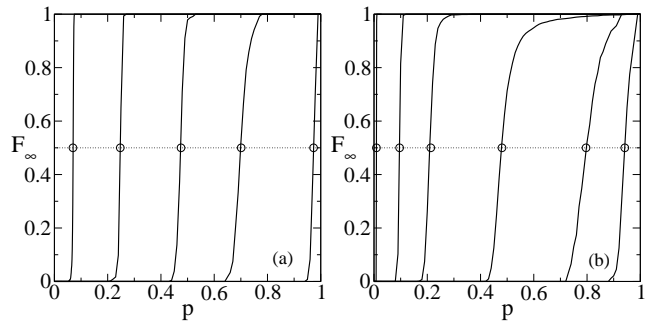}
\caption{\label{fig1} 
The ratio of non-spanning configurations vs the fraction
of removed nodes $p$. Lines are simulation data, from networks of
$N=10^6$ nodes, while the circles are the theoretical critical points.
a) Results from 100 different realizations of networks with $\gamma=2.5$.
Left to right: $\alpha$=$4$, 1, $0.5$, $0.25$ and $0$.
b) Results from 300 realizations of networks with $\gamma=3.5$.
Left to right: $\alpha$=4, 1, $0.5$, 0, $-0.5$, and $-1$.}
\end{figure}

In the numerical treatment and the analysis we use random mixing for the network construction,
where no correlations exist between the connectivity of neighboring sites.
We first construct the network for a given $\gamma$. We fix the number
of nodes $N$, as well as $m$ and assign the degree $k$ (number of links) for each node by drawing a
random number from a power law distribution $P(k)\sim k^{-\gamma}$.
We then randomly select pairs of links between nodes that have not yet 
reached their preassigned connectivity
and have not already been directly linked to each other. We repeat this selection until the
entire network has been created.

To find the percolation thresholds $p_c$, we choose successively nodes with 
probability $W(k)$ (see Eq.~\ref{Wki}) and remove them. When a node is 
removed all its links are cut. After each removal, we calculate $\av{k^2}$ 
and $\av{k}$. If $\kappa=\av{k^2}/\av{k} \ge 2$, a spanning cluster exists 
in the network. We repeat this procedure for a large number of  configurations
(typically 100-300).
For each concentration $p$ of removed nodes we determine the probability 
$F_{\infty}$ that a spanning cluster does not exist. We obtain $p_c$ from the 
condition $F_{\infty}=1/2$, as is shown in Fig.~\ref{fig1} for 
$\gamma= 2.5$ and 3.5 and several $\alpha$ values between 4 and -1.
The width of the dispersion in $F_{\infty}$ gives an upper bound for the error bars for $p_c$.

\begin{figure}
\includegraphics{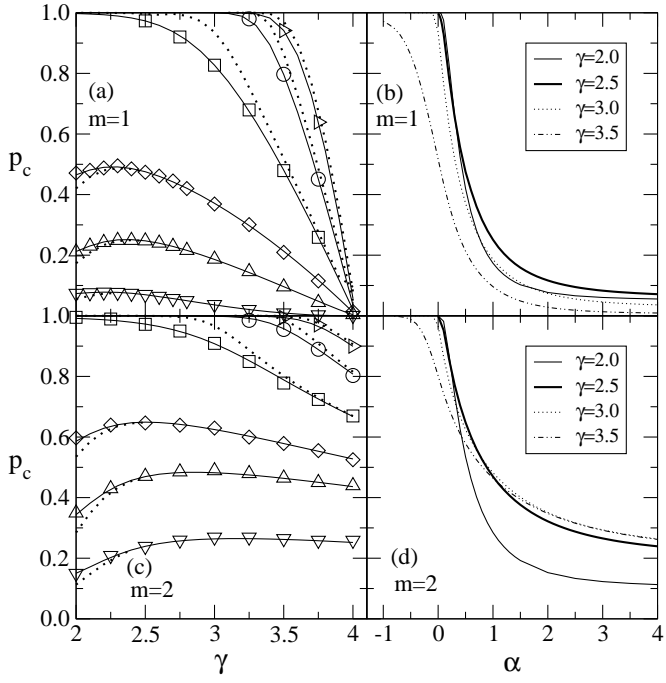}
\caption{\label{fig2} 
(a) Values of $p_c$ vs $\gamma$ for different $\alpha$ values:
(bottom to top) $\alpha=4, 1, 0.5, 0, -0.5, -1$. Symbols represent
simulation data ($N=10^6$ nodes) from $100-300$ different network
realizations. Solid lines
are the theoretical predictions for finite-size networks, while
dashed lines correspond to infinite-size networks. Lower cutoff: $m=1$,
(b) Values of $p_c$ vs $\alpha$ for infinite-size networks and different $\gamma$ values.
Lower cutoff $m=1$
(c) Same as (a), with lower cutoff: $m=2$, and
(d) Same as (b), with lower cutoff: $m=2$.}
\end{figure}

In the analytical treatment, we assume that sites are chosen for deletion
according to their initial connectivity. A site is chosen with probability
$k^\alpha/N\av{k^\alpha}$. After $d$ deletion attempts, the
probability that a site of connectivity $k$ has not been deleted is
\begin{equation}
\rho(k)=\left(1-\frac{k^\alpha}{N\av{k^\alpha}}\right)^{d}\approx
e^{-dk^\alpha/N\av{k^\alpha}}\;.
\end{equation}
 The condition for the existence of a spanning cluster after the attack 
is as follows: a site is reached through a link with probability 
$kP(k)/\av{k}$ and is still functional with probability $\rho(k)$.
If the average number of outgoing links ($k-1$) per site is larger than
$1$, a spanning cluster will exist. This consideration is formulated by:
\begin{equation}
\label{an_qc}
\sum_{k=m}^{k_{\rm max}} \frac{P(k)k(k-1)}
{\av{k}}e^{-q_c k^\alpha}=1\;,
\end{equation}
where $q\equiv d/N\av{k^\alpha}$, and $q_c$ is the value of $q$ at criticality.
To find the fraction of removed sites, one numerically solves Eq.~(\ref{an_qc})
to calculate $q_c$ and substitutes this value for calculating
the critical threshold of removed sites:
\begin{equation}
\label{an_pc}
p_c=\sum P(k)\rho(k)= \sum_{k=m}^{k_{\rm max}} P(k)\exp(-q_c k^\alpha)\;.
\end{equation}

Equation~(\ref{an_pc}) describes the percolation threshold $p_c$ for a 
given network of $N$ nodes with exponent $\gamma$, as a function of the 
attack parameter $\alpha$.

The lines in Fig.~\ref{fig2}
represent the solutions of Eqs.~(\ref{an_qc}) and (\ref{an_pc}) and are 
in excellent agreement with the simulations (in symbols). Remarkably, for $\gamma<3$, $p_c$ becomes smaller than $1$ already for very small
positive $\alpha$ values,  and decays rapidly with
increasing $\alpha$. Accordingly, by a very small preference probability to remove
highly connected nodes, which arises, for example, in an intentional attack with very little knowledge of the network structure,
this network can be destroyed by removing a comparatively small
fraction of nodes. Above $\alpha =\gamma-1$, $p_c$ saturates, which means that
the knowledge available to the attacker in this case is sufficient to destroy the network most efficiently.
From Figs. 2a and c one might conclude that
 negative values of $\alpha$ might lead to $p_c=1$ also for
$\gamma$ values above 3.
 We tested this question numerically and found 
that for all values of $\gamma$ values between 3 and 4 and negative $\alpha$, the percolation threshold  $p_c$ is below 1.
For large negative values of  $\alpha$ and $\gamma>3$, the critical threshold 
can be approximated by
$p_c=1-(\gamma-1)\left[2\left(\frac{\gamma-3}{\gamma-2}\right)
m^{2-\gamma}\right]^{\frac{\gamma-1}{\gamma-3}}$, which is below 1 for all $\gamma>3$ and $m\geq 2$.

\begin{figure}
\includegraphics{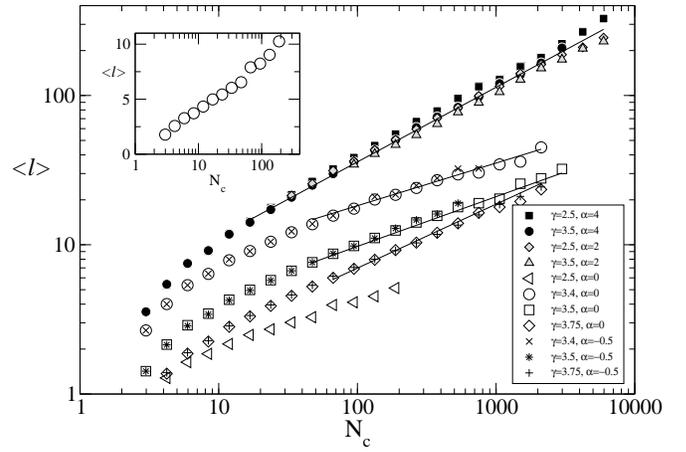}
\caption{\label{fig} 
Average shortest distance $\langle l\rangle$ between any two nodes of the
giant cluster at criticality as a function of the cluster size $N_c$.
The results correspond to networks of initially $N=10^4, 10^5$, and $10^6$
nodes. 1000 different configurations have been used for each $N$, except 
for
$N=10^6$ (100 configurations). The data have been logarithmically binned, and the results have been vertically shifted for presentation clarity.
The values of $\alpha$ and $\gamma$ are shown in the plot.
The lines represent the theoretical slopes of
(top to bottom): $1/2, 1/3.5$, $1/3$, and $1/2.33$, respectively.
Inset: Semi-logarithmic plot for the case of $\gamma=2.5$ and $\alpha=0$.}
\end{figure}

To study the effect of the different attack strategies on the topology of the network 
just before disruption, we analyzed the topology of the network at  the percolation transition for different values 
of $\gamma$ and $\alpha$. We characterize the topology  by the way the average shortest topological distance $\av{l}$ between two nodes
scales with the cluster size $N_c$. We expect that $N_c$ scales with $\av{l}$ as $N_c \sim \av{l}^{d_{\ell}}$, where $d_{\ell}$ is the topological
(``chemical'') dimension.
Using a mean-field type approximation
Ref.~\cite{Cohen02}, it has been  suggested that for random removal,
\begin{equation}
\label{eq_dl}
d_\ell = \frac{\gamma-2}{\gamma-3} \; , \;\;\; 3<\gamma<4,
\end{equation}
while for the intentional attack
\begin{equation}
\label{eq_dl_gamma2}
d_\ell =2 \; , \;\;\; \gamma>2.
\end{equation}

The power-law dependence $\av{l}\sim N_c^{1/d_{\ell}}$ is very different from
the logarithmic dependence  $\av{l}\sim \log N$ found in scale-free networks for
$p$ well above $p_c$ and $\gamma>3$. Accordingly, due to the attacks, the network becomes
very inefficient, since the distances between the nodes increase drastically, from
a logarithmic to a power-law dependence on the total number of nodes. Equations (8)
and (9)
 suggest that random removal and intentional attack are in different universality classes.
This implies that the critical properties of the percolation transition depend on the way the
transition is being approached, which is quite unusual in critical phenomena.
To test these predictions and to see if there are further universality classes,
we studied numerically how, at criticality, $\av{l}$ scales with $N_c$, for different
networks and different attack 
parameters $\alpha$.

Figure 3 shows simulation results for $\av{l}$ as a function of $N_c$ for several values of $\gamma$ and $\alpha$.
For $\alpha>0$, the data scale quite nicely and yield a slope very close to $1/2$ (corresponding to $d_l=2$) for all $\gamma$ values,
being identical to the theoretical prediction, Eq.~(\ref{eq_dl_gamma2}), for the $\alpha=\infty$. This shows  that all attacks with
$\alpha>0$ fall into the same universality class.
The figure also verifies the prediction of Eq.~(\ref{eq_dl}) for random removal ($\alpha=0$) when $\gamma>3$. It shows, in addition, that attacks with
$\alpha<0$ result almost in the same network structure as for $\alpha=0$, and yield the same topological dimension.  Thus, for a given network with $\gamma>3$
the same network can undergo transitions of two universality
classes: a) $\alpha>0 $ (universality class of the targeted intentional attack), and b)  $\alpha\leq 0$ (universality class of random removal).

For $2<\gamma<3$ and $\alpha\leq 0$, the situation is less conclusive.
Here, it is difficult to distinguish between a logarithmic or a power-law dependence,
as can be seen from the inset of Fig. 3. Since for the pure network, $\av l\sim \log\log N$
\cite{Cohen02}, also a simple logarithmic dependence $\av l\sim \log N$ will lead to a strong
increase of the mean distance between the nodes at criticality, as suggested by the inset. 

In summary, we have studied
the network tolerance under different attack 
 strategies. 
We find that little knowledge on the highly connected nodes in an intentional attack reduces the threshold drastically compared with the random case.
For example, when in a  scale-free network with $\gamma=2.5$ the one percent 
 highest connected nodes are known with  probability $w=0.2$ (corresponding to
$\alpha=1$), the 
threshold reduces from $p_c=1$ for the random case ($w=0$) to
$p_c\cong 0.25$. When all hubs are known ($w=1$), $p_c$ is close to 0.07. This shows that
for example the Internet (see also \cite{Li}) can be damaged efficiently when only a small fraction of hubs is
known to the attacker. Moreover, this result is also relevant for immunization of
populations: Even if the virus spreaders are known (and immuned) with small 
probability,  the spreading threshold can be reduced significantly.
We also showed that even if the attack does not yet disintegrate the network,
there is nevertheless a major damage on the network, since the distances
between the nodes increase significantly and any transport process
on the net may become inefficient.
Our results show that the topology of the network close to criticality, characterizing the universality class
of the phase transition, depends on the strategy of node removal.

\begin{acknowledgments}
This work was supported by a European research NEST project DYSONET 012911, by the DAAD and by the Israel Science 
Foundation.
\end{acknowledgments}


\begin{thebibliography}{}
\bibitem{footnote}
An example for a negative value of $\alpha$ is the  breakdown of computer networks,
where it can  be assumed that low degree computers are more prune to
malfunction due to hardware failure, maintenance, viruses, power outs etc.
than high degree computers. We may assume, for instance, that in a
university network, the down time of the main routers having hundreds of
connections is about 10 times less than of end units, and the hubs for
the different LANs (Local Area Networkss) are somewhere in between. In this case, the situation is
similar to an attack with $\alpha=-1/2$.
\end{thebibliography}
\end{document}